\documentclass[useAMS,usenatbib]{mn2e}
\usepackage[total={17.8cm,24.0cm},centering]{geometry}
\usepackage{graphicx}
\usepackage{color}
\usepackage{times}
\usepackage{amsmath}
\usepackage{multirow}
\title[An Oxford SWIFT IFS study of ETGs in the Coma cluster]
  {An Oxford SWIFT Integral Field Spectroscopy study of 14 early-type galaxies in the Coma cluster}
\author[N. Scott et al.]{Nicholas Scott,$^{1,2}$\thanks{E-mail: nscott@astro.swin.edu.au}, Ryan Houghton$^2$, Roger L. Davies$^2$,  Michele Cappellari$^2$,\newauthor Niranjan Thatte$^2$, Fraser Clarke$^2$, Matthias Tecza$^2$\\
$^1$Centre for Astrophysics \& Supercomputing, Swinburne University of Technology, PO Box 218, Hawthorn, VIC 3122, Australia\\
$^2$Sub-Dept. of Astrophysics, Dept. of Physics, University of Oxford, Denys Wilkinson Building, Keble Road, Oxford, OX1 3RH, UK\\}
\date{May 2012}
\pagerange{\pageref{firstpage}--\pageref{lastpage}} \pubyear{2012}

\def\LaTeX{L\kern-.36em\raise.3ex\hbox{a}\kern-.15em
    T\kern-.1667em\lower.7ex\hbox{E}\kern-.125emX}

\begin{document}

\label{firstpage}

\maketitle

\begin{abstract}
As a demonstration of the capabilities of the new Oxford SWIFT integral field spectrograph, we present  first observations for a set of 14 early-type galaxies in the core of the Coma cluster. Our data consist of {\it I-} and {\it z-}band spatially resolved spectroscopy obtained with the Oxford SWIFT spectrograph, combined with {\it r-}band photometry from the SDSS archive for 14 early-type galaxies. We derive spatially resolved kinematics for all objects from observations of the calcium triplet absorption features at $\sim 8500$ \AA. Using this kinematic information we classify galaxies as either Fast Rotators or Slow Rotators. We compare the fraction of fast and slow rotators in our sample, representing the densest environment in the nearby Universe, to results from the ATLAS$^\mathrm{3D}$ survey, finding the slow rotator fraction is $\sim 50$ per cent larger in the core of the Coma cluster than in the volume-limited ATLAS$^\mathrm{3D}$ sample, a 1.2 $\sigma$ increase given our selection criteria. Comparing our sample to the Virgo cluster core only (which is 24 times less dense than the Coma core) we find no evidence of an increase in the slow rotator fraction. Combining measurements of the effective velocity dispersion $\sigma_e$ with the photometric data we determine the Fundamental Plane for our sample of galaxies. We find the use of the average velocity dispersion within 1 effective radius, $\sigma_e$, reduces the residuals by 13 per cent with respect to comparable studies using central velocity dispersions, consistent with other recent integral field Fundamental Plane determinations.
\end{abstract}


\section{Introduction}
In the local Universe, integral field spectroscopy surveys have shed new light on the formation and evolution of nearby early-type galaxies (ETGs). In particular, two dimensional information has led to a significant revision of the way we classify these objects. ETGs had long been known to exhibit a range of kinematic types, from the traditional picture of giant, pressure-supported elliptical galaxies \citep{Binney:1982} to less massive, rotationally supported systems \citep{Davies:1983}. The SAURON \citep{deZeeuw:2002} and ATLAS$^\mathrm{3D}$ \citep{Cappellari:2011a} surveys improved this classification system to take advantage of integral field spectroscopy, basing their classification on the observed specific stellar angular momentum of a galaxy, $\lambda_R$ \citep{Emsellem:2007}. \citet{Emsellem:2011} classified galaxies into fast-rotators (FRs, $\lambda_R/\sqrt{\epsilon} > 0.31$) and slow-rotators (SRs, $\lambda_R/\sqrt{\epsilon} \leq 0.31$, where $\epsilon$ is the galaxy's ellipticity). This classification system differs significantly from the traditional E/S0 morphological classification, with the majority of early-types being FRs and only a small fraction being SRs (14 per cent in the ATLAS$^\mathrm{3D}$ survey, compared to 27 per cent being classified `elliptical'). This classification scheme provides an interesting new perspective on the morphology-density relation \citep{Dressler:1980}.  Dividing galaxies into spirals, FRs and SRs, \citet{Cappellari:2011b} find that the fraction of all galaxies that are FRs increases linearly with environment, in the sense that the densest environments have a higher proportion of FRs. In contrast they find that the fraction of all galaxies that are SRs (and correspondingly the fraction of ETGs that are SRs) increases only in the densest environment covered by the ATLAS$^\mathrm{3D}$ survey, the core of the Virgo cluster. By determining the fraction of ETGs that are SRs (N$_\mathrm{SR}$/N$_\mathrm{ETG}$, hereafter the SR fraction) in the Coma cluster, the densest environment in the nearby Universe, we can extend the kinematic morphology-density relation to denser environments and gain insight into which physical processes give rise to this relation.

\begin{table*}
\caption{Photometric and spectroscopic parameters of the SWIFT IFS Coma ETG sample}
\label{tab:sample}
\begin{center}
\begin{tabular}{l c c c c c c c c}
\hline
Name & Exposure & R$_e$ & $\langle \mu_{e,r} \rangle$ & $\epsilon$ & $\sigma_e$ & Fractional IFS & $\lambda_\mathrm{R}$ & Classification\\
& Time (s) & (arcsec) & (mag arcsec$^{-2}$) & & (km\ s$^{-1}$) & Coverage & &\\
(1) & (2) & (3) & (4) & (5) & (6) & (7) & (8) & (9) \\
\hline
\hline
PGC44679 & 3600 & 4.07 & 18.81 & 0.14 & $119 \pm 6$ & 1.0 & 0.11& SR\\
IC4011 & 3600 & 5.41 & 19.12 & 0.07 & $105 \pm 8$ & 0.58 & 0.28 & FR\\
IC4051 & 2400 & 17.67 & 19.93 & 0.19 & $231 \pm 16$ & 0.46 & 0.04 & SR\\
NGC4860 & 1800 & 6.99 & 18.27 & 0.15 & $213 \pm 9$ & 1.0 & 0.10 & SR\\
NGC4867 & 2400 & 2.35 & 16.98 & 0.30 & $183 \pm 10$ & 1.0 & 0.20 & FR\\
NGC4872 & 1800 & 4.35 & 18.05 & 0.23 & $213 \pm 14$ & 1.0 & 0.34 & FR\\
NGC4873 & 3600 & 8.57 & 19.24 & 0.21 & $153 \pm 11$ & 1.0 & 0.17 & FR\\
NGC4874 & 4800 & 54.31 & 20.78 & 0.07 & $200 \pm 26$ & 0.31 & 0.08 & SR\\
NGC4883 & 2700 & 7.49 & 18.99 & 0.17 & $149 \pm 11$ & 1.0 & 0.15 & FR\\
NGC4886 & 3600 & 7.16 & 18.93 & 0.03 & $128 \pm 11$ & 1.0 & 0.06 & FR\\
NGC4889 & 1800 & 31.61 & 19.45 & 0.30 & $375 \pm 32$ & 0.56 & 0.11 & SR\\
PGC44533 & 3600 & 4.95 & 19.66 & 0.23 & $94 \pm 5$ & 0.92 & 0.20 & FR\\
PGC44602$^*$ & 3600 & 6.00 & 19.78 & 0.36 & $79 \pm 5$ & 0.75 & 0.21 & FR\\
PGC44662 & 2400 & 4.46 & 18.68 & 0.28 & $127 \pm 5$ & 0.99 & 0.22 & FR\\
\hline
\end{tabular}
\end{center}
$^*$ Observed with the 0.16 arcsec pixel scale\\
Columns: (1) Galaxy name from Hyperleda. (3) Effective radius, typical error 10 per cent. (4) Average surface brightness within one R$_e$, typical error 10 per cent. (5) Ellipticity measured at 1 R$_e$. (6) Effective velocity dispersion. (7) Fraction of 1 R$_e$ covered by the SWIFT IFS field-of-view. (8) $\lambda$ parameter measured at 1 R$_e$. Typical errors on $\lambda_\mathrm{R}$ are $\sim 5$ per cent. (9) Kinematic classification as \citet{Emsellem:2011}: SR = Slow Rotator, FR = Fast Rotator.
\end{table*}

Perhaps the key tool in understanding and interpreting ETG structure is the Fundamental Plane (FP): the observed relationship between the effective radius, R$_e$, effective surface brightness, $\langle \mu \rangle_e$ and velocity dispersion ($\sigma$) of a galaxy. The tight observed relationship between these quantities, first noted by \citet{Djorgovski:1987} and \citet{Dressler:1987}, has been used to interpret the structure and evolution of ETGs. This is possible because the Virial Theorem predicts a FP relationship for relaxed, dynamically hot stellar systems \citep[see e.g.][]{Binney:1998}. The observed FP is found to be tilted relative to the Virial Theorem prediction, as well as showing small but non-zero intrinsic scatter. The tilt and intrinsic scatter of the observed FP contain significant information about the structure of ETGs \citep[see][for a summary of some possible physical origins of the FP tilt]{Renzini:1993}. An IFS study of ETGs in a single cluster brings two key improvements to the currently available FP observations: i) by selecting only cluster members we select galaxies at a single common distance, eliminating errors in distance as a source of observed scatter in the FP and ii) most studies of the FP have relied upon measurements of the central velocity dispersion $\sigma_c$, whereas the Virial Theorem prediction for the FP relates to global variables; specifically, $\sigma_\mathrm{global}$ which is better represented by $\sigma_e$, which can only be measured using integral field spectroscopy. 

In a future work we will present dynamical models of all galaxies in our sample (for which IFS data is a key ingredient), allowing us to construct the mass plane \citep[relating galaxy mass to $\sigma_e$ and R$_e$ --][]{Bolton:2007} for ETGs in the Coma cluster. We will also use stellar population modelling to derive metallicities from the calcium triplet and use these to explore correlations between the stellar populations, dynamics and metallicity gradients of the galaxies in our sample.

In Section \ref{Sec:Sample} we present our Coma ETG sample including details of how it was selected and outline the observations. In Section \ref{Sec:Data} we discuss the measurement of the photometric properties of our sample galaxies, the reduction of the SWIFT IFS data and the measurement of the kinematic properties. We then compare the results of our analysis to that of the ATLAS$^\mathrm{3D}$ survey in Section \ref{Sec:Comparison}. In Section \ref{sec:FP} we briefly present the Fundamental Plane for the initial SWIFT IFS Coma sample.

\section{Sample and observations}
\label{Sec:Sample}
\subsection{Sample}
The sample of this SWIFT IFS Coma study consists of 14 ETGs, selected from the catalogue of \citet{Scodeggio:1998}. Our galaxies lie within 0.25 degrees ($\sim 0.2$ Mpc) of the centre of the Coma cluster \citep[RA $=12^{\rm h}59^{\rm m}36^{\rm s}$, DEC $=+27^{\circ} 58^{\prime}$][]{Hammer:2010}. \citet{Sparke:2007} define the region within 0.2 Mpc of the cluster centre as the core region and in their table 7.1 give a density for this region of 13,429 galaxies Mpc$^{-3}$, the densest region in the local Universe. We divided the \citet{Scodeggio:1998} catalogue into seven logarithmically spaced bins in velocity dispersion. From each bin we selected two galaxies, giving reasonable sampling of the range of velocity dispersions and masses found in ETGs in the Coma cluster. Properties of the current SWIFT IFS sample are given in Table \ref{tab:sample}. 

\subsection{IFS Observations}
Our IFS observations were carried out using the Short Wavelength Integral Field specTrograph \citep[SWIFT,][]{Thatte:2006}, an integral field spectrograph based on the image slicer concept with wavelength coverage from 6500 to 10500 \AA. The SWIFT IFS is mounted on the Cassegrain focus of the 200" Hale Telescope at the Palomar Observatory in California. The observations were carried out over three separate observing runs (3rd/4th of May 2009, 25th/26th March 2010, 5th June 2010), using the largest 0.235 arcsec pixel$^{-1}$ spatial scale with a field-of-view of 10.3 x 20.9 arcsec$^2$. For the three largest galaxies (IC4051, NGC4874 and NGC4889) we mosaiced multiple pointings to increase the IFS coverage towards 1 R$_e$. The seeing varyied between 0.9 and 2 arcseconds, with a median value of 1.3 arcseconds.

\begin{figure*}
\centering
\includegraphics[width=1.5in,clip,trim=25 10 35 0]{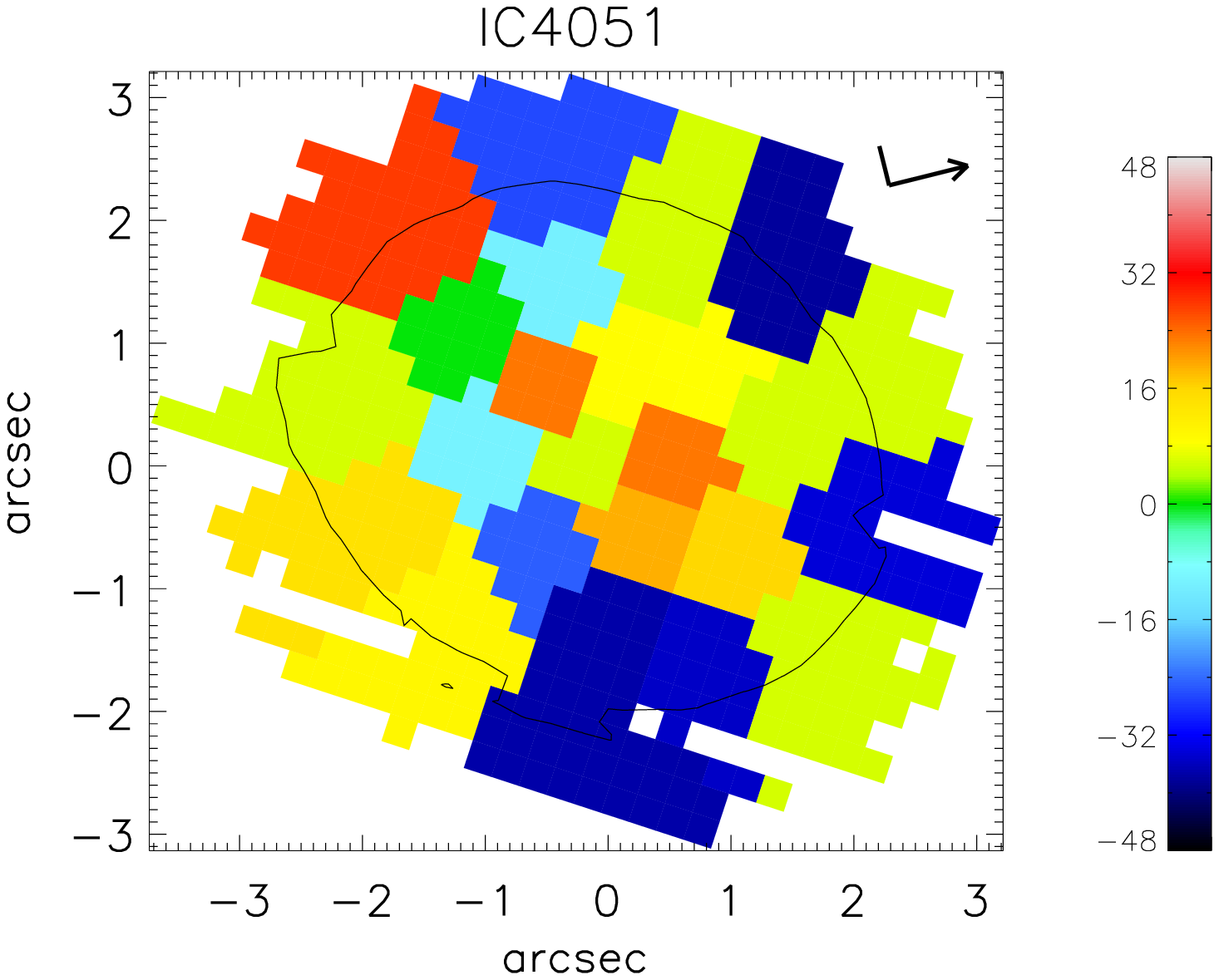}
\includegraphics[width=1.5in,clip,trim=15 10 75 0]{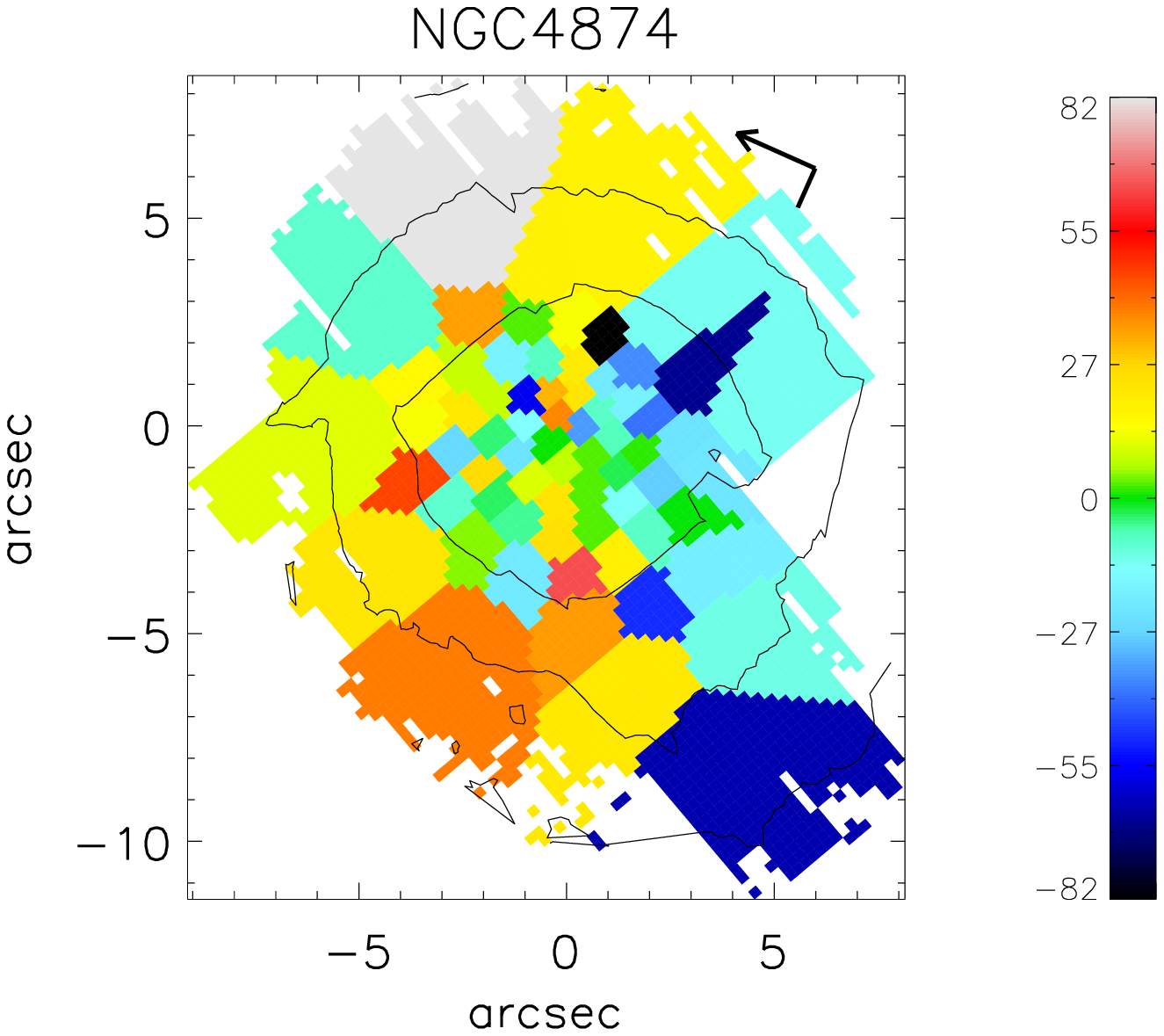}
\includegraphics[width=1.5in,clip,trim=25 10 35 5]{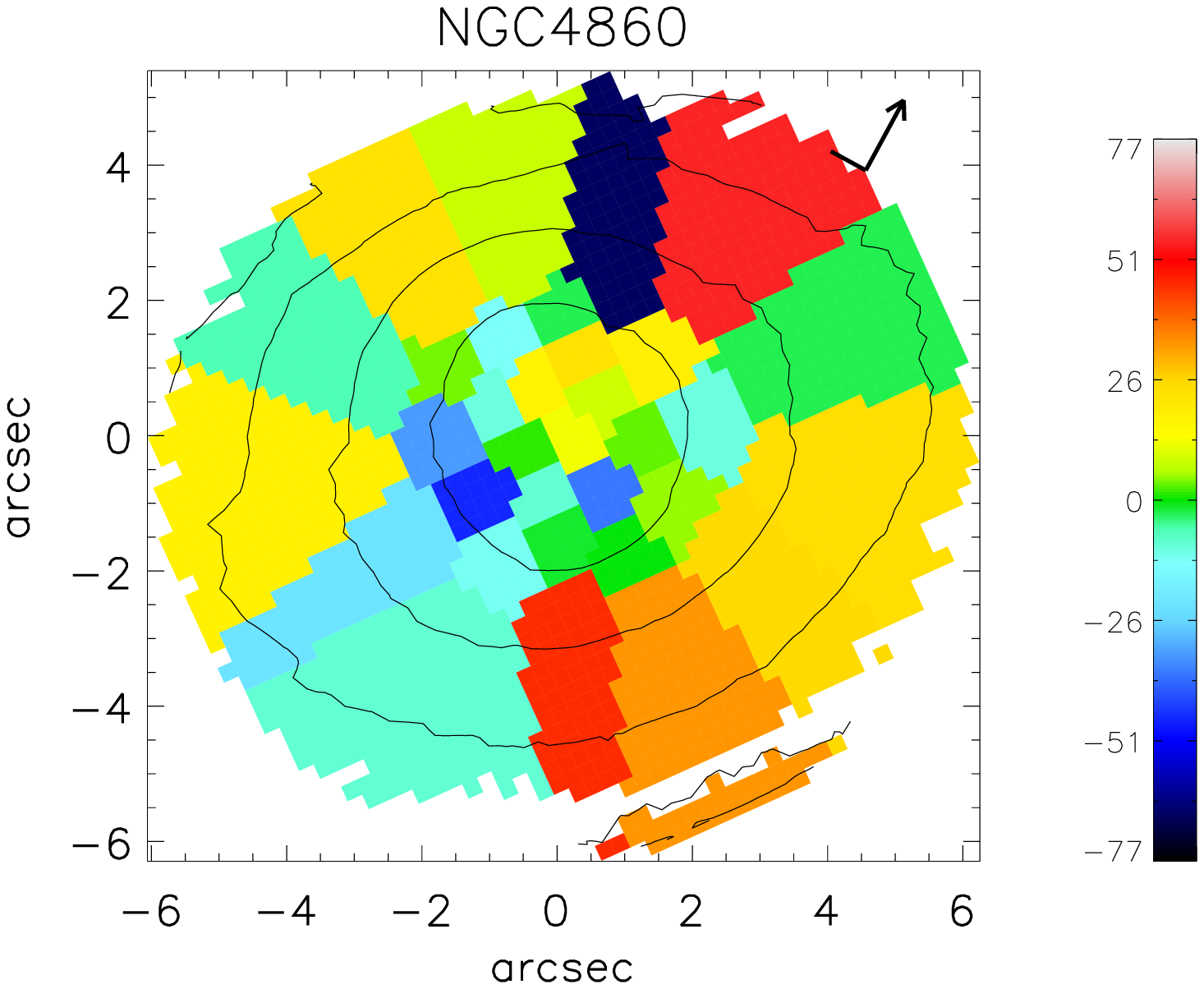}
\includegraphics[width=1.5in,clip,trim=25 10 35 5]{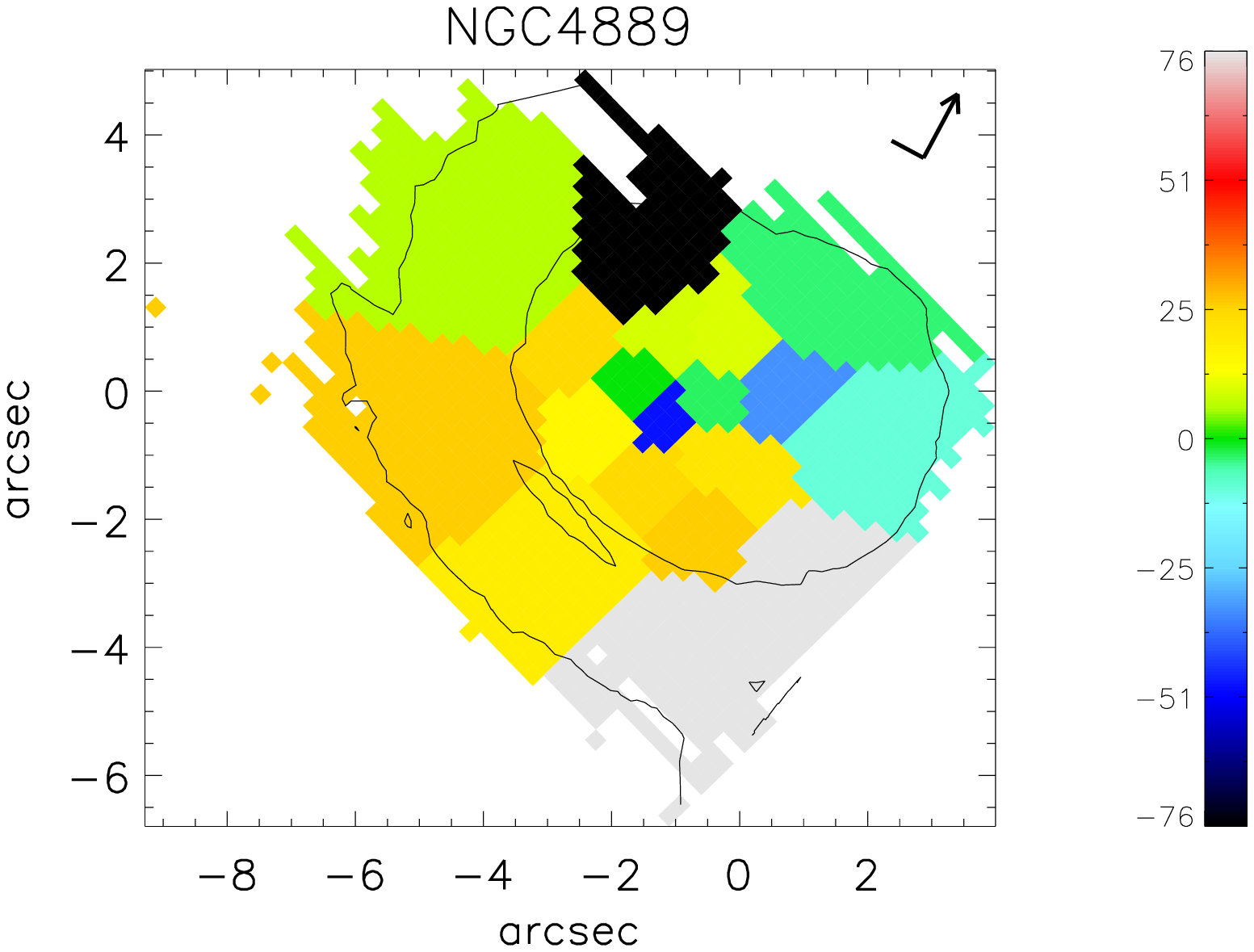}
\includegraphics[width=1.5in,clip,trim=25 10 35 5]{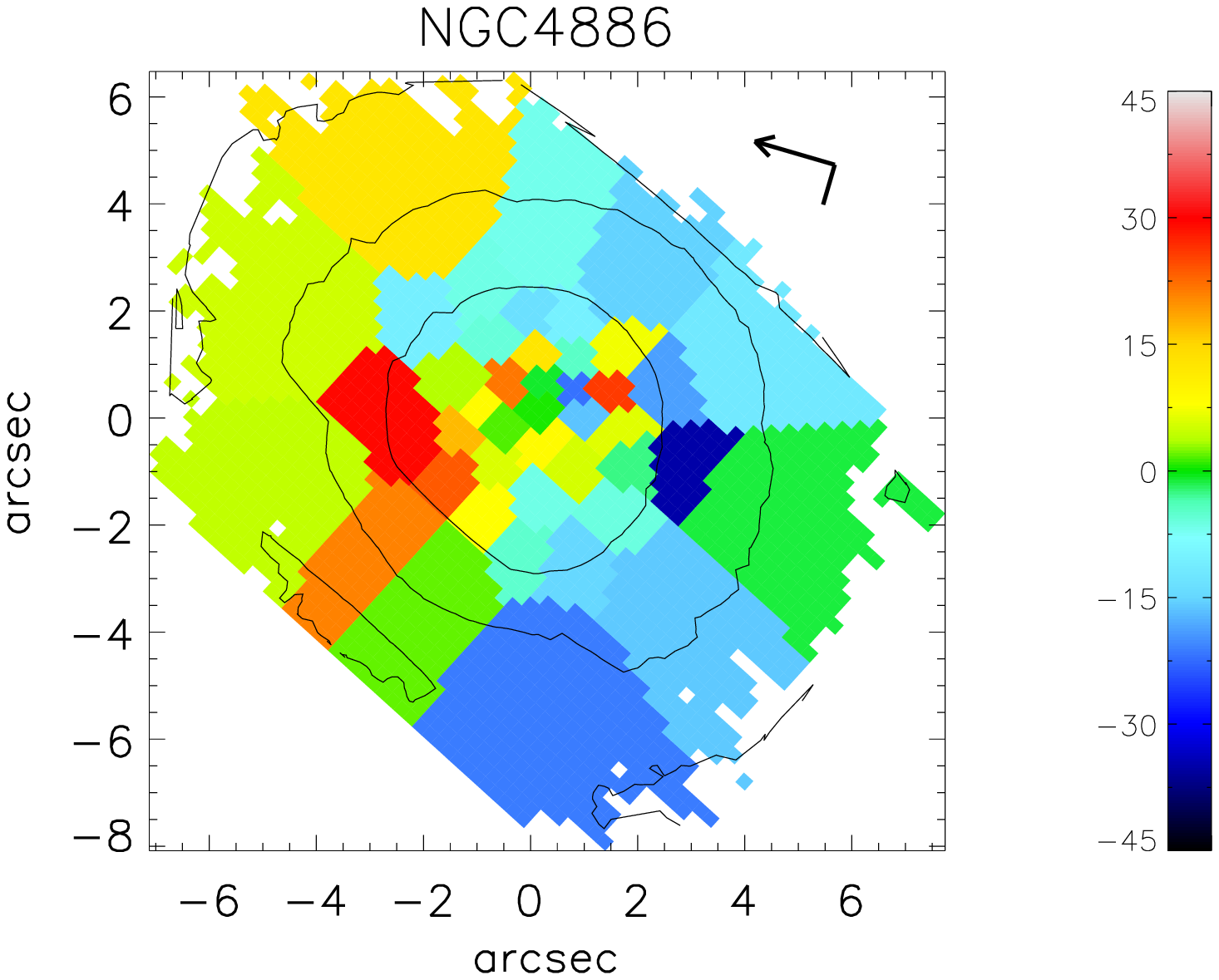}
\includegraphics[width=1.5in,clip,trim=25 10 25 5]{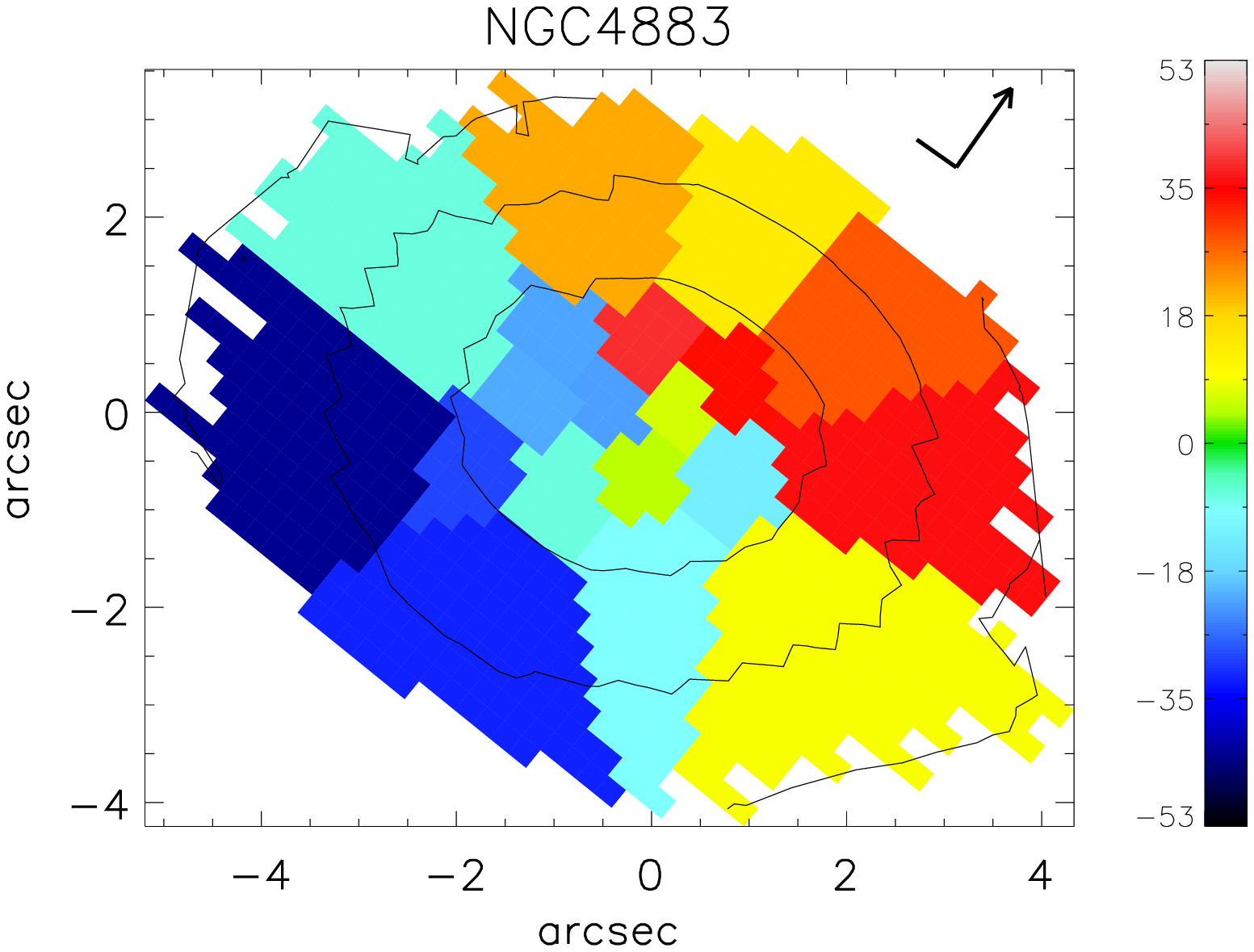}
\includegraphics[width=1.5in,clip,trim=25 10 25 0]{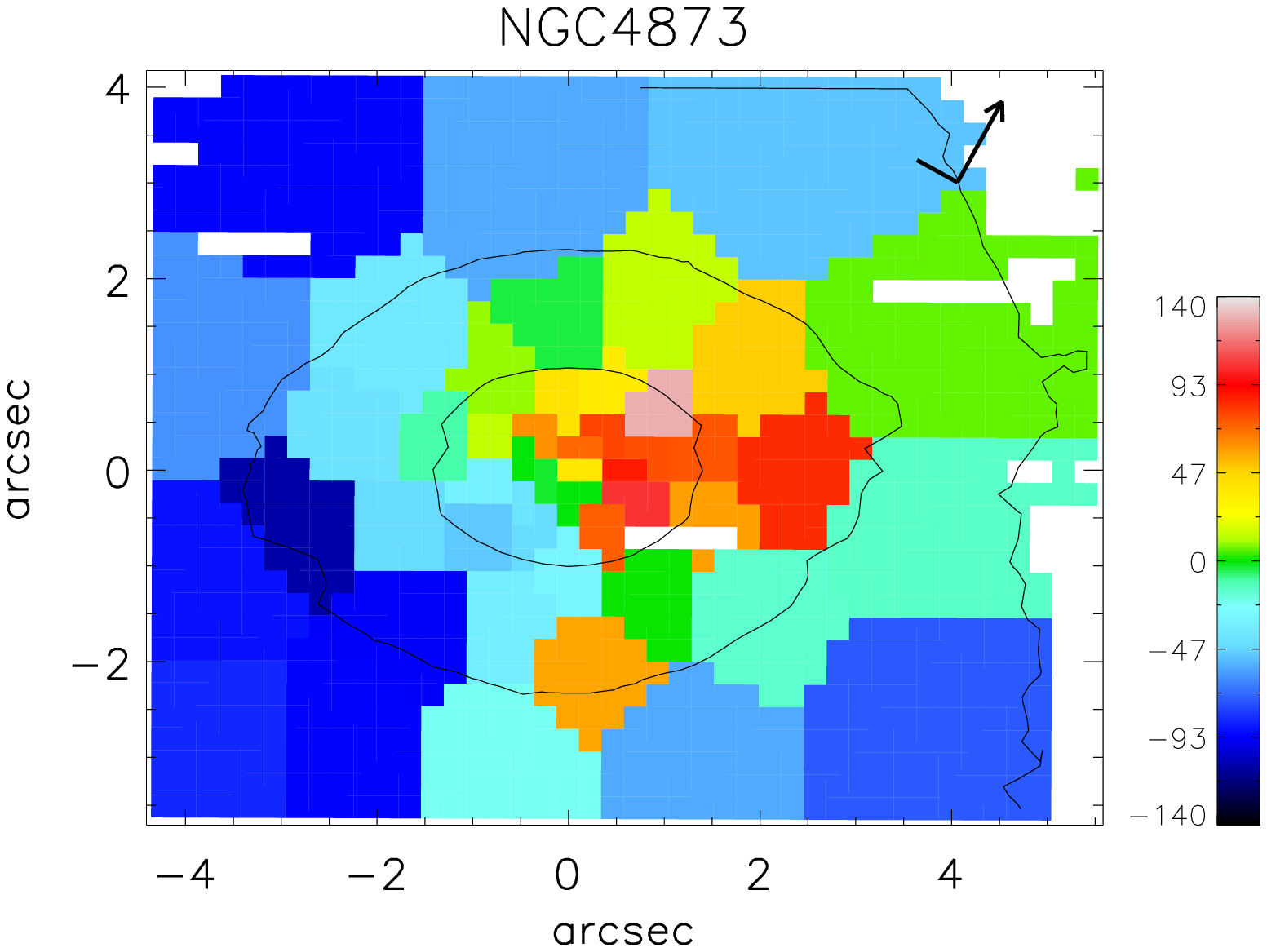}
\includegraphics[width=1.5in,clip,trim=25 10 55 0]{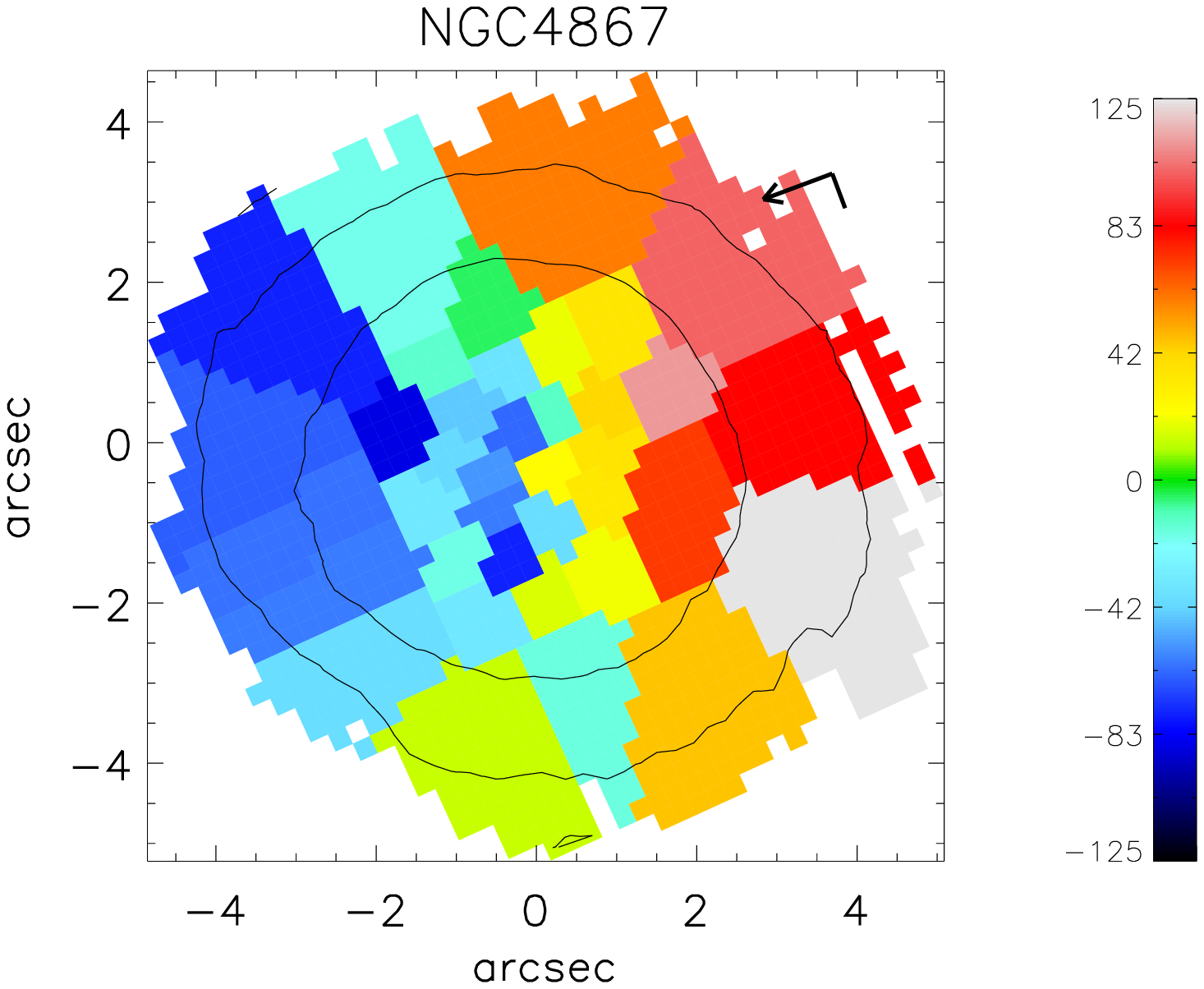}
\includegraphics[width=1.5in,clip,trim=25 10 35 0]{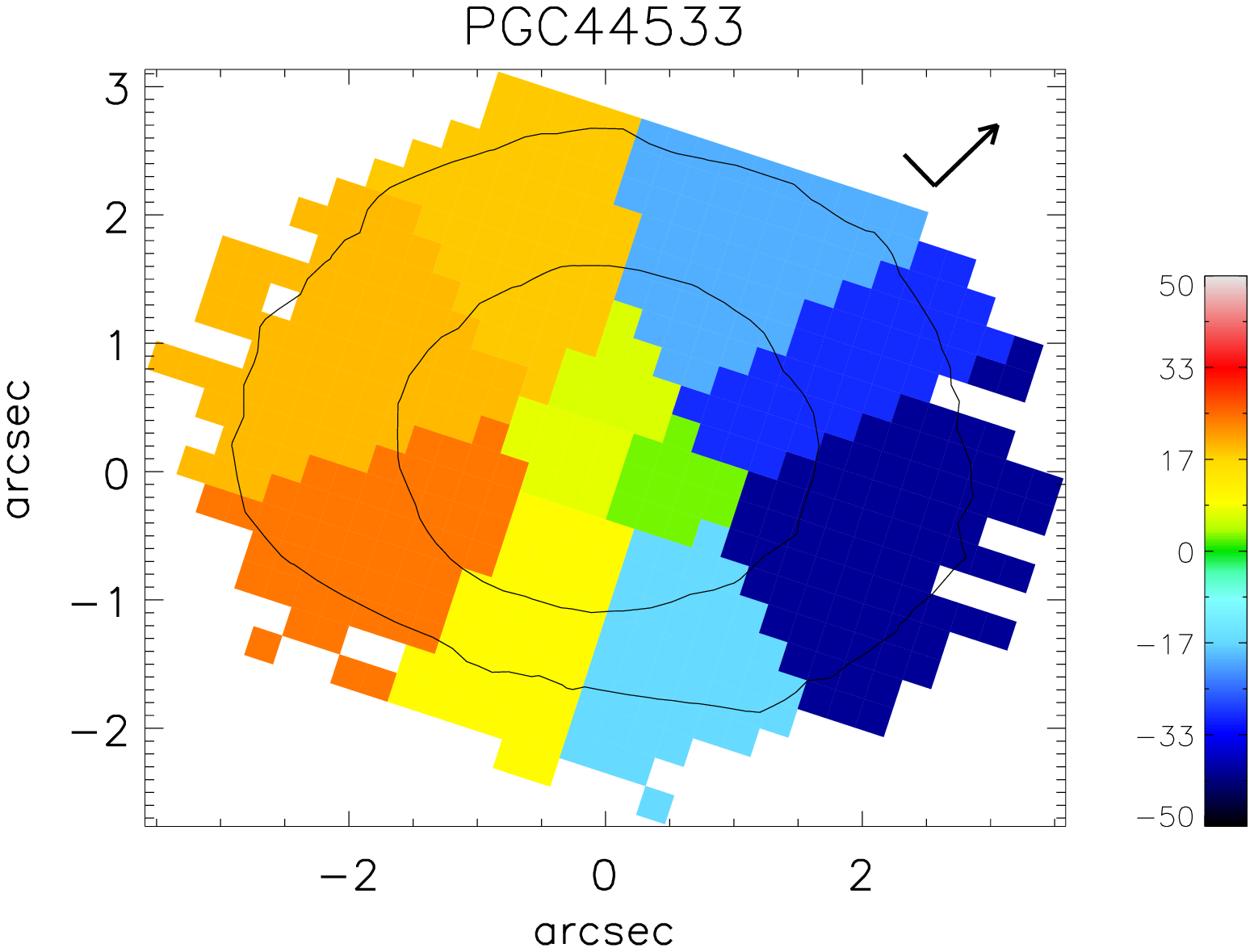}
\includegraphics[width=1.5in,clip,trim=25 10 35 75]{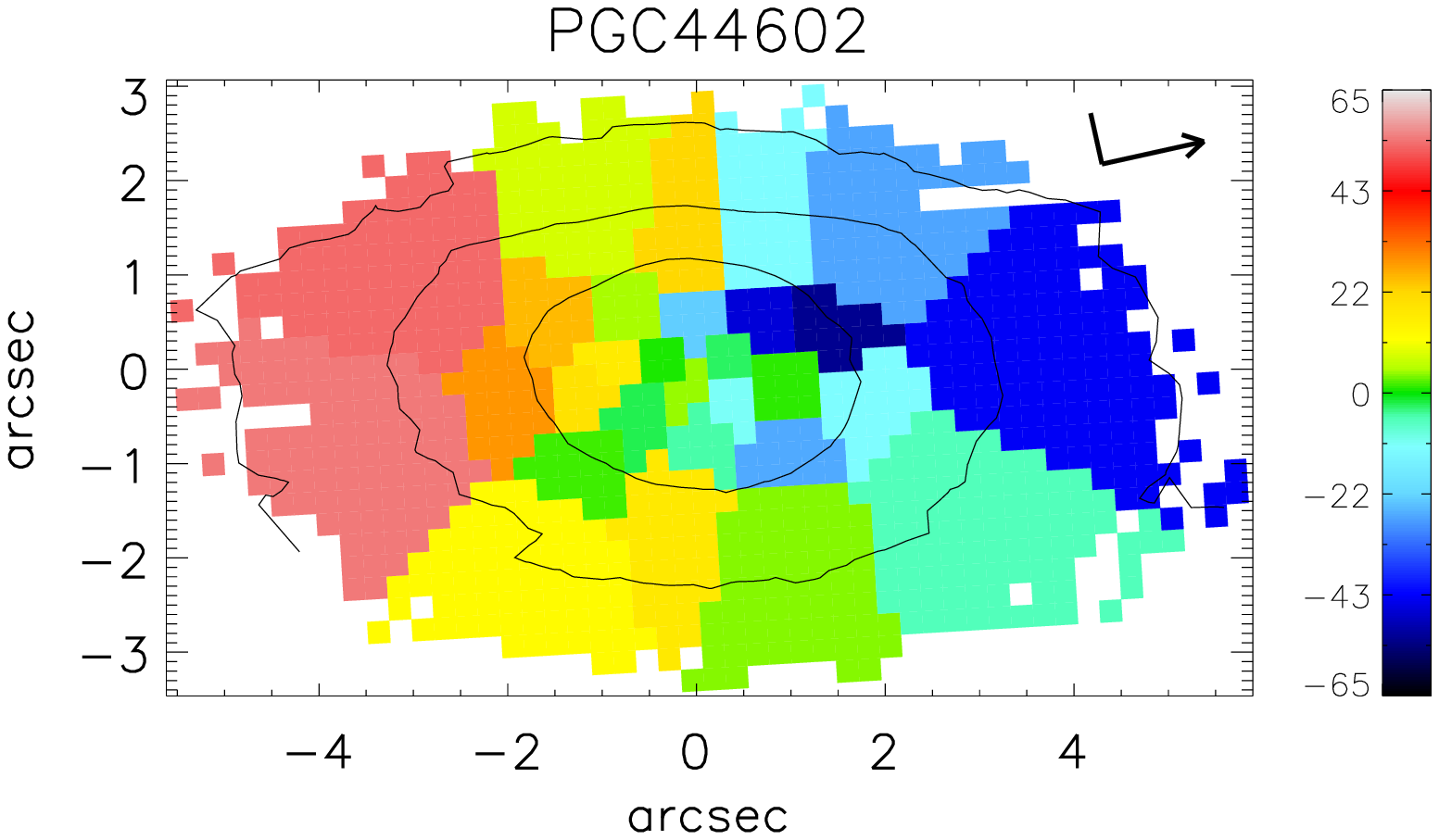}
\includegraphics[width=1.5in,clip,trim=25 10 35 5]{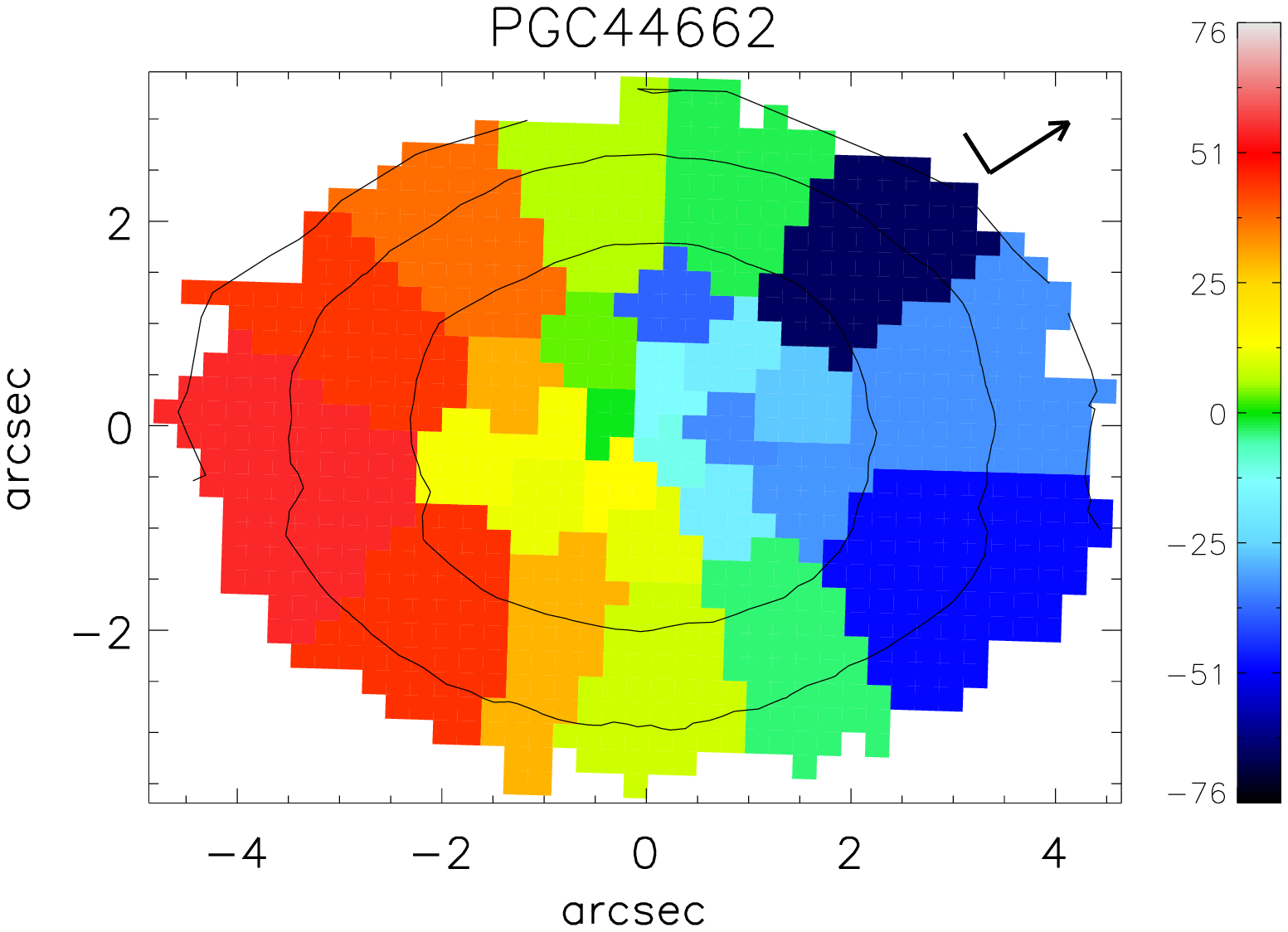}
\includegraphics[width=1.5in,clip,trim=25 10 65 0]{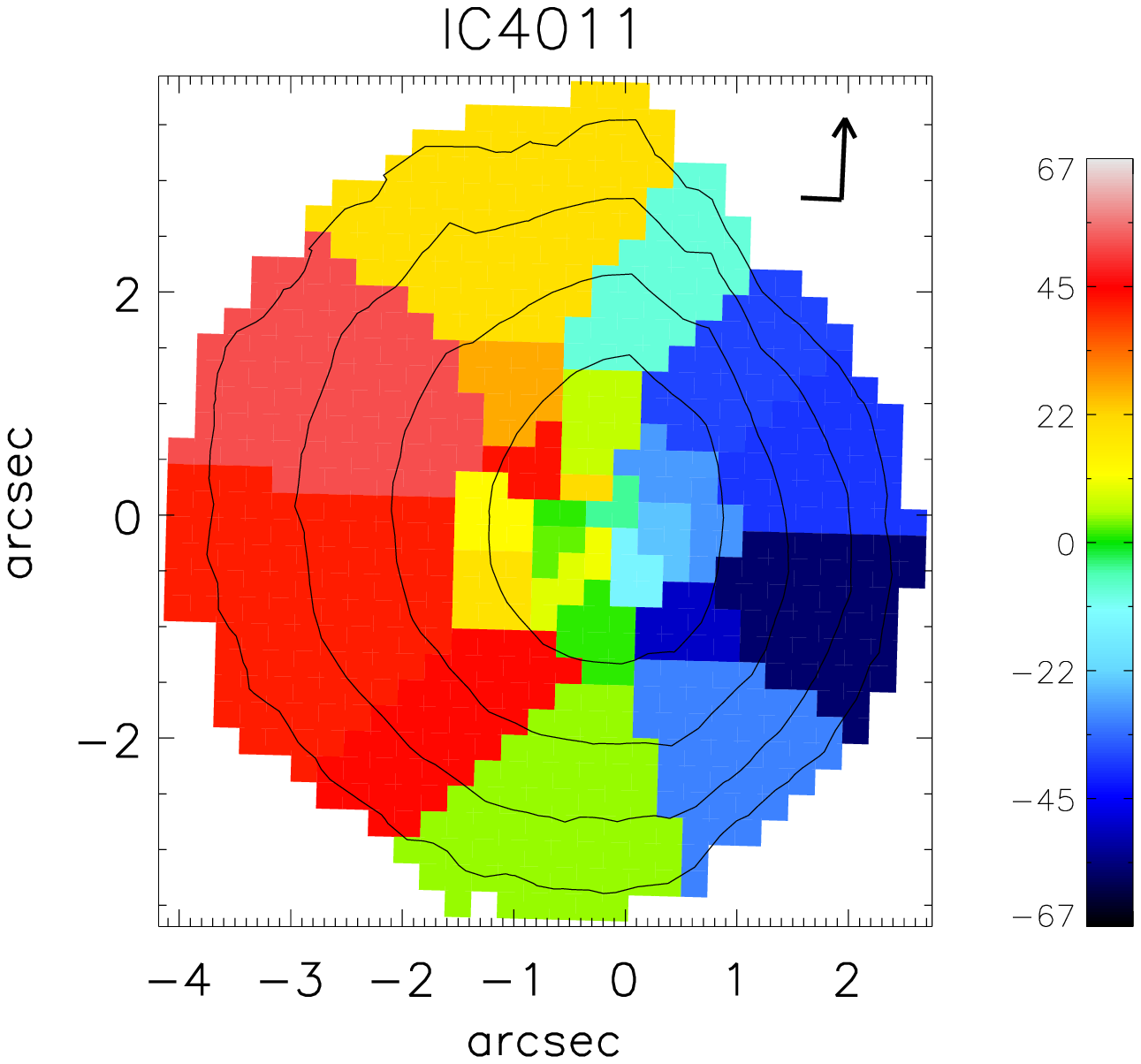}
\includegraphics[width=1.5in,clip,trim=25 10 35 0]{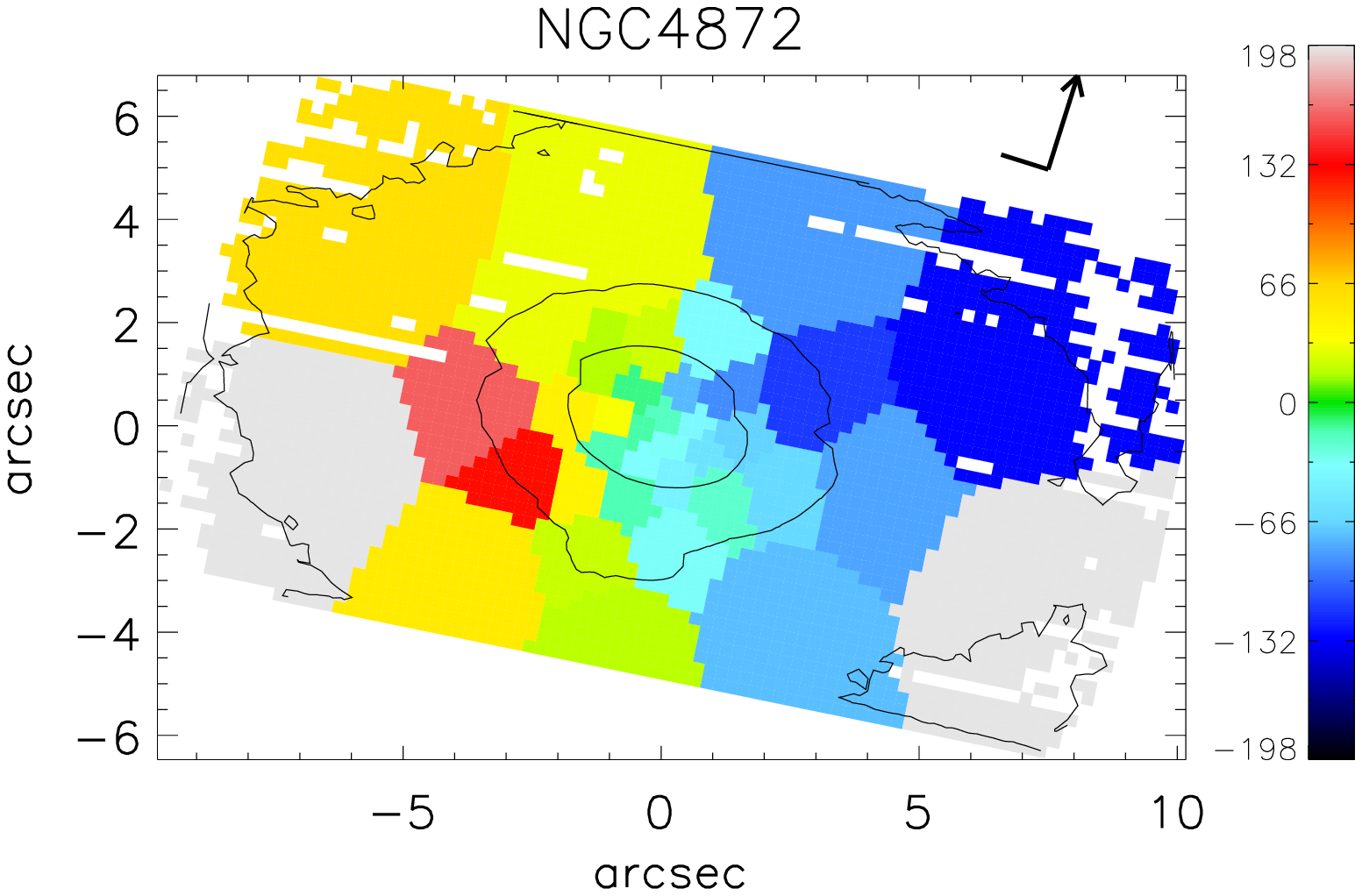}
\caption{Velocity maps for 13 Coma ETGs in the SWIFT IFS sample. The top row shows the 4 SRs, the remaining rows show the FRs. The range of velocities in each map is given by the adjacent colour bar. Representative isophotes from the SWIFT data are overplotted. The maps are rotated such that the photometric position angle is aligned with the $x$ axis. The N--E direction is indicated with the arrow and dash in the upper right corner of each panel.}
\label{fig:velocity_fields}
\end{figure*}

\subsection{Imaging}
Optical {\it r-}band images for the full sample were obtained from the Sloan Digital Sky Survey \citep[SDSS,][]{York:2000} Data Release 7 \citep{Abazajian:2009}, via the Montage \citep{Berriman:2004} web service\footnote{\it http://hachi.ipac.caltech.edu:8080/montage}, which automatically mosaics, flux calibrates and background subtracts individual SDSS fields. The final images have a spatial scale of 0.4 arcsec pixel$^{-1}$ and were obtained in median seeing of 1.4 arcseconds. 

\section{Data Analysis}
\label{Sec:Data}
\subsection{Photometric quantities}
Photometric quantities ($\epsilon$, R$_e$ and $\langle \mu_{e,r} \rangle$) were measured from the SDSS {\it r-}band images. We made used the curve of growth technique \citep[e.g.][]{Cappellari:2006} to determine R$_e$ and $\langle \mu_{e,r} \rangle$ directly from the sky-subtracted images. We determined the ellipticity from the moment technique described in \citet{Cappellari:2007}. Using the {IDL} routine {\it find$\_$galaxy}\footnote{Available as part of the MGE package of \citet{Cappellari:2002}.}, the second moments of the luminosity distribution were determined for all pixels belonging to the galaxy above a fixed brightness, the level being chosen such that the pixels selected encompass a region with radius 1 R$_e$ or matching the extent of the IFS observations, whichever is the smaller. The ellipticity was then derived from the observed second moments of the luminosity distribution, which is directly comparable to the measurements presented in the ATLAS$^\mathrm{3D}$ survey \citep{Emsellem:2011}. The photometric quantities are given in Table \ref{tab:sample}.

\subsection{Spectroscopic quantities}
\subsubsection{Data Reduction}
The SWIFT IFS data were reduced using a dedicated {\sc IRAF}-based pipeline, {\it swiftred} (Houghton et al., in preparation). The pipeline involves the standard data reduction steps of bias subtraction, flat-fielding, illumination correction, image reconstruction and wavelength calibration. In addition, cosmic rays were removed using the LACOSMIC routine of \citet{vanDokkum:2001}, implemented in the {\it swiftred} package. We make use of the OH night sky emission lines \citep[using the emission line atlas of][]{Hanuschik:2003} in the science observations to provide an additional wavelength calibration and spatial flexure correction data. The final wavelength calibration is accurate to $\sim 0.1$ \AA. The sky spectrum was removed by subtracting blank sky frames adjacent in time to each science exposure. The subtraction of sky emission lines was accurate to $\sim 2$ per cent of the emission line flux.

\subsubsection{Kinematics}
We extract the stellar kinematics using the penalised PiXel Fitting (pPXF)\footnote{Available from http://purl.org/cappellari/idl} technique of \citet{Cappellari:2004}. pPXF uses a library of stellar spectra to produce a template spectrum and convolves this with a line of sight velocity distribution (LOSVD) to best match the observations. We constrain the LOSVD to have a Gaussian shape -- we do not fit for the higher-order Gauss-Hermite moments. For fitting the CaT absorption features we used the stellar spectra library of \citet{Cenarro:2001}, which contains $\sim 700$ spectra of stars covering a broad range in age and metallicity. Any residuals from cosmic rays or inaccurate sky subtraction were cleaned at this stage by rejecting data points with residuals $> 3 \sigma$ from the model spectrum (using the keyword {\sc /CLEAN} in the pPXF software) and repeating the fit. Only a single rejection iteration was required to produce stable measurements. Large sky residuals were masked from the fitting process.

For galaxies where the IFS field of view was $> 2$ R$_e$, an R$_e$ aperture spectrum was formed by binning all pixels lying within an elliptical aperture with area $\pi \rm{R}_e^2$ and ellipticity $\epsilon_e$. For galaxies larger than this an aperture spectrum was formed by accreting spaxels into an aperture of ellipticity $\epsilon_e$ until the S/N dropped below 2 per spectral pixel. At this point the diminishing returns from adding additional pixels makes little difference to the quality of the aperture spectrum or the accuracy of the measured $\sigma_e$. The fraction of R$_e$ covered by this aperture spectrum is given in Column 7 of Table \ref{tab:sample}. We measure the mean recession velocity $V$ and effective velocity dispersion $\sigma_e$ from these aperture spectra (Table \ref{tab:sample}). For those galaxies where our integrated aperture does not extend to $1 R_e$ we correct the measured dispersion according to Eqn. (1) from \citet{Cappellari:2006}. The magnitude of this correction is small -- for NGC4874 (the galaxy with the smallest fractional coverage of R$_e$) it is only 8 per cent. 

To measure $\lambda_R$ for our galaxies we require spatially resolved kinematics. As the object flux (hence the S/N) decreases rapidly with radius it was necessary to spatially bin the data to determine reliable kinematics at large radii. We made use of the Voronoi binning technique of \citet{Cappellari:2003} to achieve this. A S/N threshold of 40 was selected as the best compromise between extracting robust kinematics from each bin and giving good spatial coverage of each galaxy. Spaxels with a S/N $< 2$ were excluded from the Voronoi binning. pPXF was again used to determine $V$ and $\sigma$ for each bin, using the template spectrum determined from the previous fit to the R$_e$ aperture spectrum. This procedure generated V and $\sigma$ maps for each galaxy in our sample; the V maps are shown in Figure \ref{fig:velocity_fields}.

\begin{figure}
\includegraphics[height=3.25in,clip,trim = 30 15 20 30]{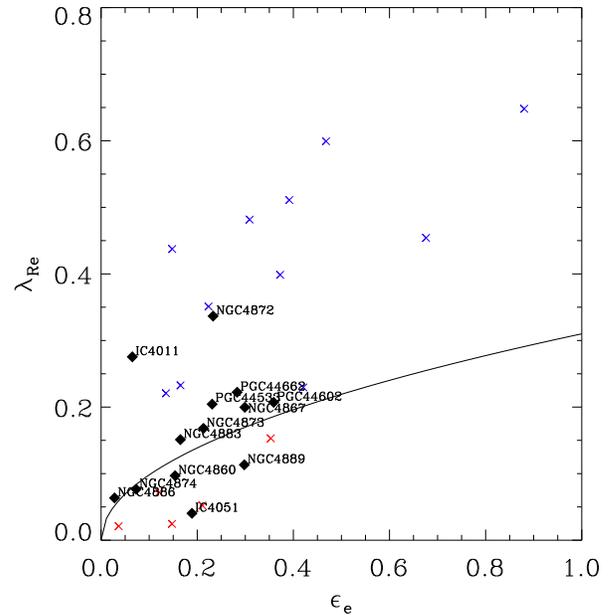}
\caption{The $\lambda_{\mathrm{R}_e}$ - $\epsilon$ diagram for the SWIFT IFS Coma ETG sample. The solid black line shows the divide between SRs and FRs given by $0.31\sqrt{\epsilon}$. 4/13 galaxies in our sample are SRs. We exclude PGC044679  from our analysis due to poor quality IFS data. The red and blue crosses indicate the position SRs and FRs respectively from the core of the Virgo cluster from \citet{Emsellem:2011}.}
\label{fig:lambdar_eps}
\end{figure}

\section{Environmental dependence of Slow and Fast Rotators}
\label{Sec:Comparison}
From the $V$ and $\sigma$ maps described in the previous section we determine $\lambda_R$ for each of the galaxies in our  sample using Eq. 6 of \citet{Emsellem:2007}. In Figure \ref{fig:lambdar_eps} we show the distribution of our galaxies in the $\lambda_\mathrm{R} - \epsilon$ plane. The solid black line marks the divide between SRs and FRs. The result of our measurements of $\lambda_\mathrm{R}$ and the corresponding FR/SR classification is given in Table \ref{tab:sample}. For PGC044679  the data quality was too poor to reliably determine $\lambda_R$. Excluding PGC044679,  we find 4 SRs and 9 FRs in the SWIFT IFS Coma ETG sample. This gives a SR fraction of 4/13 or $31 \pm 15$ per cent. NGC4886, despite having a low value of $\lambda_{\mathrm{R}_e}$, is extremely round ($\epsilon = 0.03$) and so likely to be very close to face on.

The Coma cluster is the densest environment in the local Universe, hence is an ideal place to search for environmental effects on galaxy properties. We make use of the ATLAS$^\mathrm{3D}$ survey \citep{Cappellari:2011a} to provide comparison data in lower density environments. We divide the ATLAS$^\mathrm{3D}$ sample into Virgo and non-Virgo galaxies, representing intermediate and low density environments respectively. The dependence of kinematic type on local environment for the ATLAS$^\mathrm{3D}$ sample is presented in \citet{Cappellari:2011b}. The principal drawback of our Coma ETG sample is that it is not volume limited - we only observe a fraction of the total ETG population of the Coma cluster. Fortunately the ATLAS$^\mathrm{3D}$ survey is a volume-limited survey, which allows us to ask the question: ``If we apply our sample selection criteria to the ATLAS$^\mathrm{3D}$ sample, what is the probability that we reproduce the SR fraction and distribution we find for our Coma sample?"

Using $\sigma_e$ for the ATLAS$^\mathrm{3D}$ sample obtained from the SAURON data (Cappellari, private communication) we use a Monte Carlo technique to extract samples of galaxies according to our Coma sample selection criteria from both the Virgo and non-Virgo ATLAS$^\mathrm{3D}$ samples. For each mock sample we determine the SR fraction to estimate the likelihood of `observing' a particular SR fraction given the actual SR fraction, the $\sigma$ distribution and our sample selection criteria. 

\begin{figure}
\includegraphics[width=3.25in,clip,trim = 40 0 20 25]{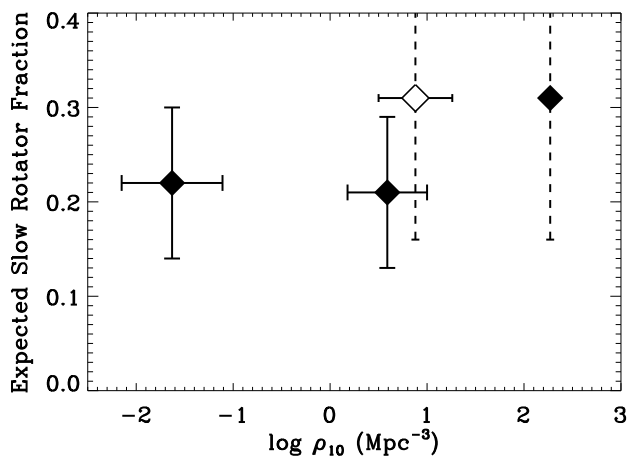}
\caption{The dependence of the slow rotator fraction on the local environment. The first two solid points show the Monte Carlo expectation from the non-Virgo and Virgo ATLAS$^\mathrm{3D}$ subsamples, given the SWIFT IFS Coma sample selection. The third solid point is from the SWIFT IFS Coma ETG sample. For the Coma datapoint no error is shown on environmental density due to the more limited sample size and selection. The error on the SR fraction is a Poisson error only and is not appropriate to compare to the ATLAS$^\mathrm{3D}$ data points. The open data point shows the {\it observed} (not matched to our sample selection criteria) slow rotator fraction in the core of the Virgo cluster from the ATLAS$^\mathrm{3D}$ survey. Note the marginal increase in the slow rotator fraction in both cluster cores compared to lower density environments but the lack of an increase in the much denser Coma core compared to the Virgo core.}
\label{fig:SR_frac_density}
\end{figure}

In Figure \ref{fig:SR_frac_density} we plot the Monte Carlo expectation of the SR fraction (given our sample selection criteria) for each environment against the projected local environmental density. Densities for the Virgo and non-Virgo ATLAS$^\mathrm{3D}$ samples are taken from \citet{Cappellari:2011b} and the density of the Coma cluster core is determined as described in the following paragraph. We use the $\rho_{10}$ density estimator, defined as the number density of galaxies in a sphere enclosing the ten nearest neighbour objects. We find little difference in the Monte Carlo expectation of the SR fraction for the ATLAS$^\mathrm{3D}$ Virgo and field samples. This is consistent with \citet{Cappellari:2011b} who, for their complete, volume-limited samples, found that the mean SR fraction in Virgo is only 16 per cent, similar to the field. Our SR fraction of $31 \pm 15$ per cent represents an increase of 1.2 $\sigma$ over the ATLAS$^\mathrm{3D}$ Virgo and field sub-samples.  

Because our sample is drawn from the core of the Coma cluster we also compare our SR fraction to that found in the core of the Virgo cluster by the ATLAS$^\mathrm{3D}$ survey. \citet{Sparke:2007} define the core of the Virgo cluster to be the region within 0.4 Mpc of the cluster centre \citep[RA $=12^{\rm h}28^{\rm m}19^{\rm s}$, DEC $=+12^{\circ} 40^\prime$][]{Mould:2000}. In their table 7.1 they give the density of this region as 559 galaxies Mpc$^{-3}$, 24 times less than that in the core of the Coma cluster. The ATLAS$^\mathrm{3D}$ core Virgo region consists of 16 galaxies -- too small a sample to match to our Coma selection criteria using the method described above -- so we compare the observed slow rotator fractions only. \citet{Cappellari:2011b} find 5/16 galaxies in the Virgo core are SRs, $31 \pm 14$ per cent. We find no increase of the slow rotator fraction in the core of the Coma cluster relative to the core of the Virgo cluster, despite a factor 24 increase in the local environmental density. 

We note that the galaxies we identify as FRs are, on average, found at lower $\lambda_R$ than those in the Virgo cluster from the ATLAS$^\mathrm{3D}$ survey (compare the blue crosses to the black datapoints in Figure \ref{fig:lambdar_eps}). The mean $\lambda_R$ for FRs in our sample is $0.21\pm0.07$; for the ATLAS$^\mathrm{3D}$ Virgo FRs over the range in $\epsilon$ covered by our sample the value is $0.33 \pm 0.14$. While our FRs in the Coma cluster are offset in $\lambda_R$, the two sets of FRs are consistent within the errors. In our sample we are missing high $\lambda_R (> 0.6)$ FRs frequently found in the ATLAS$^\mathrm{3D}$ sample. This is likely due to our sample missing very flattened galaxies. This lack of flattened systems may be a property of the Coma cluster itself -- the $\epsilon$ distribution of Coma ETGs in the \citet{Scodeggio:1998} catalogue our sample is drawn from has more low $\epsilon$ galaxies than the ATLAS$^\mathrm{3D}$ sample -- or it may be due to the unclear nature of the \citet{Scodeggio:1998} sample selection. A larger IFS sample is required to determine if the FR population of the Coma cluster is consistent with that of the ATLAS$^\mathrm{3D}$ survey.

\section{The Fundamental Plane of Coma early-type galaxies}
\label{sec:FP}

\begin{figure}
\includegraphics[width=3.25in,clip,trim = 30 10 15 25]{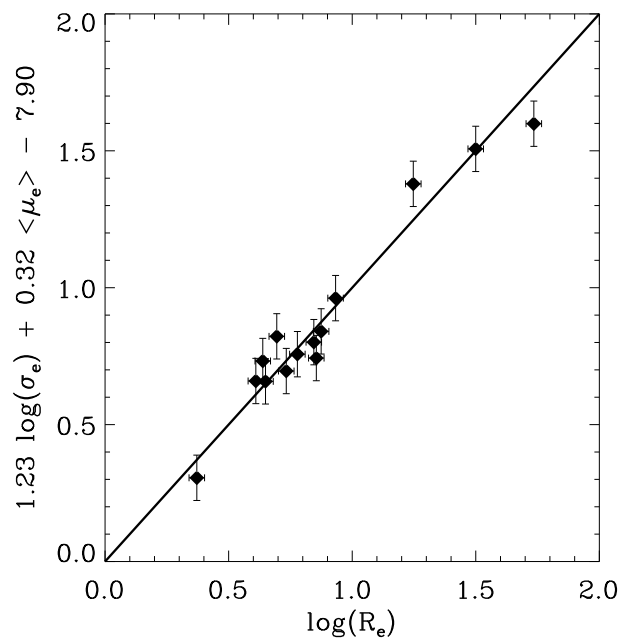}
\caption{The FP determined from the SWIFT IFS Coma ETG sample determined by minimising the residuals orthogonal to the plane. The error on $\sigma_e$ was determined by pPXF (typically $7.5$ per cent) and the R$_e$ and $\langle \mu_e \rangle$ errors are based on the results of growth curve fits with different levels of sky subtraction. (typically 10 per cent in R$_e$ and$\langle \mu_e \rangle$).}
\label{fig:SWIFT IFS_FP}
\end{figure}

We delay a detailed discussion of the FP for ETGs in the Coma cluster, using our enlarged final sample, to a future paper; here we simply present the best-fitting FP for our initial sample. We determine the FP by minimising the residuals orthogonal to the plane, as described in \citet{Jorgensen:1996}. This gives a FP of the form:
\begin{equation*}
\log R_e = (1.23 \pm 0.13) \log \sigma_e + (0.32 \pm 0.03) \log \langle \mu \rangle_e  - c
\end{equation*}
The errors are determined by a bootstrap procedure and the rms residuals in $\log R_e$ are 0.08. This projection of the data is shown in Figure \ref{fig:SWIFT IFS_FP}. \citet{Jorgensen:1996} determined the Coma FP from a sample of 79 ETGs with ground-based photometry and central $\sigma_c$ (measured in a fixed physical aperture of 0.87 kpc) for each galaxy. \citet{Thomas:2011} also determined the Coma FP from a sample of 16 ETGs but made use of composite HST and ground-based photometry and multiple long slit apertures for each galaxy. The FP coefficients we find are consistent with those studies (and others from the literature). However, our use of $\sigma_e$ is an improvement over the traditional $\sigma_c$: \citet{Cappellari:2006} show that, for a given $\sigma_c$, $\sigma_e$ varies by up to 30 per cent. For similar sample sizes, our FP residuals are $\sim 13$ per cent smaller than those of \citet{Thomas:2011}, consistent with the $\sim 15$ per cent reduction in scatter found by \citet{Falcon-Barroso:2011} when moving from $\sigma_c$ to $\sigma_e$. Detailed dynamical modelling of our IFS data would be desirable to recover dynamical masses and M/Ls and explore the mass fundamental plane \citep{Thomas:2011}.

\section{Discussion}
\label{Sec:Discussion}
In Section \ref{Sec:Comparison} we presented tentative evidence that the fraction of SRs (relative to the number of FRs) is enhanced in the core of the Coma cluster, in comparison to both the less-massive Virgo cluster and the field population. This is consistent with the findings of \citet{Cappellari:2011b}, that the SR fraction increases only in the dense core of the Virgo cluster and not in the less dense outskirts of the cluster.  This suggests that whatever mechanism forms SRs operates efficiently only in high density environments. Perhaps more significantly, we do not observe an increase in the SR fraction in the core of the Coma cluster relative to the core of the Virgo cluster, despite a factor 24 difference in the local density between the two environments. This suggests that whatever mechanism is responsible for the increase in the SR fraction does not depend in a simple way on the local environmental density.

To form a slow rotator it is likely that two ingredients are required: i) to remove (or cut off) the cold gas supply and ii) some form of dynamical interaction to remove the angular momentum of the galaxy. The hot intergalactic medium found in cluster environments can be effective both in ram-pressure stripping existing gas from a galaxy \citep{Giovanelli:1985,Morganti:2006} and in preventing the acquisition of new gas by cold accretion \citep{Oosterloo:2010} or recycling of stellar mass loss \citep{Leitner:2011}. While interactions are more common in dense environments these interactions tend to occur at high relative velocities, hence the dynamical effects are small \citep[but see][]{Tillson:2011}. \citet{Moore:1999} suggest that dynamical pre-processing in in-falling groups enhances the historical number of low-velocity interactions for galaxies in massive clusters. This suggests a scenario where galaxies have their total angular momentum reduced through numerous major and/or minor mergers as they fall into a massive cluster, where the increased density of the hot IGM inhibits the regrowth of a high angular momentum component. It is unclear whether there is a critical threshold in the IGM density which prevents the reacquisition of angular momentum by a galaxy -- we plan to obtain IFS observations of a larger sample of galaxies to be able to draw firm conclusions.

\section{Acknowledgements}
NS acknowledges support of Australian Research Council grant DP110103509. MC acknowledges support from a Royal Society University Research Fellowship.\\
The Oxford SWIFT Integral Field Spectrograph is directly supported by a Marie Curie Excellence Grant from the European Commission (MEXT-CT-2003-002792, Team Leader: N Thatte). It is also supported by additional funds from the University of Oxford Physics Department and the John Fell OUP research Fund. Additional funds to host and support SWIFT at the 200-inch Hale Telescope on Palomar are provided by Caltech Optical Observations.\\
Based on observations obtained at the Hale Telescope, Palomar Observatory, as part of a collaborative agreement between the California Institute of Technology, its divisions Caltech Optical Observatories and the Jet Propulsion Laboratory (operated for NASA) and Cornell University.\\
This research made use of Montage, funded by NASA's Earth Science Technology Office, Computational Technologies Project, under Cooperative Agreement Number NCC5-626 between NASA and the California Institute of Technology. The code is maintained by the NASA/IPAC Infrared Science Archive.\\
\bibliographystyle{mn2e}
\bibliography{swift_coma}

\label{lastpage}

\end{document}